# Hot-Carrier Separation in Heterostructure Nanowires observed by Electron-Beam Induced Current

**Jonatan Fast,[1] Enrique Barrigon,[1] Mukesh Kumar,[1] Yang Chen, [1] Lars Samuelson,[1] Magnus Borgström,[1] Anders Gustafsson, [1] Steven Limpert, [1] Adam Burke,[1] Heiner Linke[1]**

[1] NanoLund and Solid State Physics, Lund University, Box 118, 22100 Lund, Sweden

E-mail: Jonatan.fast@ftf.lth.se, heiner.linke@ftf.lth.se

**Abstract**

The separation of hot carriers in semiconductors is of interest for applications such as thermovoltaic photodetection and third-generation photovoltaics. Semiconductor nanowires offer several potential advantages for effective hot-carrier separation such as: a high degree of control and flexibility in heterostructure-based band engineering, increased hot-carrier temperatures compared to bulk, and a geometry well suited for local control of light absorption. Indeed, InAs nanowires with a short InP energy barrier have been observed to produce electric power under global illumination, with an open-circuit voltage exceeding the Shockley-Queisser limit. To understand this behaviour in more detail, it is necessary to maintain control over the precise location of electron-hole pair-generation in the nanowire. In this work we perform electron-beam induced current measurements with high spatial resolution, and demonstrate the role of the InP barrier in extracting energetic electrons.We interprete the results in terms of hot-carrier separation, and extract estimates of the hot carriers' mean free path.

## 1. Introduction

Hot-carrier photovoltaic (HCPV) devices are designed to separate photo-excited, hot electron-hole pairs on a timescale faster than carrier thermalisation (roughly 1-10 ps in bulk semiconductors [1]). By preventing thermalisation losses, photovoltaic conversion efficiencies as high as 66% can in theory be reached [2]. Such a device requires an energy filter through which hot carriers can be extracted at energies above the bandgap of the absorber before they relax in energy. Due to the short timescale available for carrier extraction, the fabrication of such energy filters demand an ability to perform band-engineering at the nanoscale. In many cases, nanostructuring has also been used with the aim of increasing the hot-carrier life time [3–5]. Semiconductor nanowires are a promising platform for HCPV devices for three main reasons: 1) nanowires provide freedom in band gap engineering with atomically sharp heterostructure interfaces [6] without concerns of lattice-mismatch [7]; 2) reducing the nanowire diameter increases the hot carrier temperatures, possibly by the formation of a phonon bottleneck [8,9]; 3) the geometry is well suited for controlling the location of light absorption.

Previous work [10,11] has realized InAs (small bandgap) single-nanowire HCPV devices, by implementing a short InP (large bandgap) segment, resulting in a potential barrier [12,13] that serves as an energy filter. The proposed mechanism of current generation is described in figure 1(a). By generating electron-hole pairs in a smaller region within the nanowire, the resulting variations in hot carrier concentration will drive a diffusion of carriers in both directions along the nanowire. Depending on their initial energy and relaxation time, carriers have a chance of diffusing across the barrier before thermalising with the lattice (figure 1(a)), eventually trapping them on the other side of the barrier. In zinc blende InAs the effective mass is expected to be greater for holes than electrons, with a ratio $m_e/m_h$ on the order of 0.1 at the band edges (for transport along the nanowire).[14] During photoexcitation, holes will thus receive a smaller portion of the excitation energy than electrons and be less likely to cross the barrier, resulting in a separation of charges that may be used to drive a current and produce power.

In previous studies [10,11], such devices have been observed to yield a photocurrent response under global illumination of the entire device. To confirm the mechanism proposed in Fig. 1a, and to rule out alternative explanations such as current generation at the contacts, it is desirable to control the location of electron-hole pair generation relative to the position of the barrier. This would allow for more direct observation of the barrier's role in separating charges and investigate the typical distances within which hot carriers can diffuse and reach the barrier.

By use of electron-beam induced current (EBIC) [15], a beam of highly energetic electrons (kV range) is focused onto the sample while simultaneously detecting any resulting current in the material, in our case a nanowire (see figure 1(b)). The incoming electrons deposit their energy via a cascade of inelastic scattering events that excite electron-hole pairs, where one single incoming high-energy electron may excite on the order of $10^3$ electron-hole pairs [15]. These excited carriers are initially presumed to not be in thermal equilibrium with the lattice and considered as hot carriers. The beam's spot size can be very small (order of a few nm), but the resolution is limited by the larger excitation volume resulting from secondary electrons spreading out as they scatter through the sample, which may vary on the scale of nm to μm depending on acceleration energy and material [15]. If the sample contains some mechanism whereby electron-hole pairs are separated from each other, a resulting current can be detected without applying any biasing voltage. EBIC has previously been employed to study nanowires containing pn-junctions[16–19], nanowire-metal Schottky contacts [20], and nanowires containing different types of heterostructures [21,22].

In this work, we employ EBIC to locally excite carriers in single InAs nanowire devices containing a single, axial InP–barrier. The results confirm that the barrier embedded in the nanowire separates energetic electrons from holes, in agreement with the model for current generation proposed in figure 1(a). Further, our results yield an electron diffusion length in InAs nanowires on the order of 100 nm, which provides valuable information for the design of future, optimized devices.

## 2. Method

The nanowires used for this study were grown using chemical-beam epitaxy (CBE) from 40 nm diameter Au aerosol seed particles deposited on InAs (111)B substrates. The wurtzite (WZ) InAs nanowires are roughly 2 μm long, and a 25 nm long WZ InP segment was grown in the center of the nanowires (see inset of figure 1(b)). The length of the InP segment is chosen so that tunneling through the barrier is not expected. Device fabrication was conducted as detailed in Ref. [23,24]. Nanowires were mechanically deposited on a Si substrate covered with a 100 nm thick $SiO_2$ layer for electrical insulation. The nanowires were contacted to pre-defined gold pads by one step of electron-beam lithography (EBL) followed by metal evaporation of 25 nm Ni followed by 100 nm Au. Here we present results from four devices, all of which were fabricated simultaneously and with nanowires from the same growth run to ensure uniform results.

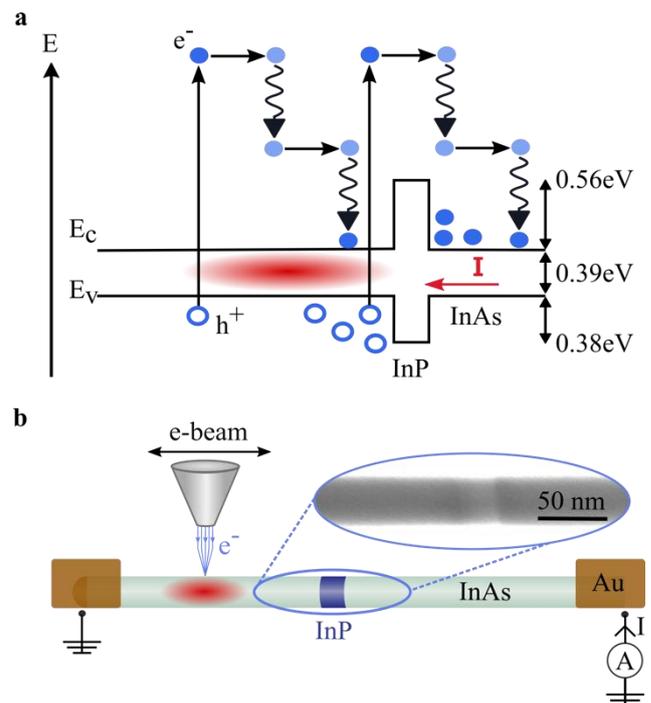

**Figure 1.** (a) An InP segment in an InAs NW represents a potential barrier for both electrons and holes [12,13]. Carriers generated on one side of the barrier may diffuse over the barrier or relax to the band edge before they reach the barrier. If the electrons have higher kinetic energy than the holes (as expected in InAs), mostly electrons will make it across the barrier, separating electron-hole pairs. (b) A contacted single NW is illuminated by an electron beam and current through the nanowire is recorded as a function of beam location (EBIC). Inset shows a scanning transmission electron microscope image of the barrier region.

Additional devices, including from other growth runs, were also studied with qualitatively consistent results.

The EBIC measurements were performed at room temperature inside a Hitachi Su8010 scanning electron microscope (SEM), operating at an acceleration voltage of 3 kV and a probe current of about 20 pA. For electrical measurements, two tungsten nano-probes with piezoelectronic positioning (Kleindiek Nanotechnik) were used to make electrical contact to each gold pad at each side of the NW device. One contact was grounded and EBIC current was measured at the other contact using a current amplifier- and measurement- system from Point Electronic. No external bias voltage was applied across the nanowire. As the electron-beam was scanned over the sample, secondary electrons were detected to create a standard scanning electron microscope (SEM) image. Simultaneously, the short-circuit current through the nanowire was measured and mapped to the position of the rastering electron beam (figure 1(b)).

The EBIC data was smoothed to reduce noise likely caused by charging of the insulating $SiO_2$ substrate, as previously observed in EBIC measurements [21]. The smoothing was done by subtracting the median background signal along each line scan (see supplemental information for details).

## 3. Results

Figure 2(a) shows the results of the EBIC measurement from one device as a composite SEM image with EBIC data overlaid. The red and blue contrast scale represents EBIC data with positive and negative current direction, respectively. We note that if the contacts are swapped such that current is measured on the opposite side of the device, the EBIC reverses sign as well. Based on TEM images of nanowires from the same growth, the InP segment is expected to be located roughly in the center of the nanowire, where we also observe the change of polarity of the current.

Figure 2(b) shows a line profile of the EBIC signal along the nanowire in-between the Au contacts. The maximal current detected is around 1 nA which is on the order of 100 times greater than the probe current. This ratio provides strong evidence that the majority of the current originates from electron-hole pairs generated by the electron beam, whereas any contribution of the probe current itself is expected to be no more than 1% of the signal. Additionally, if no hot carriers crossed the barrier, it would be expected to act as a current divider (blocking the flow of thermalized carriers) with no current detected on the grounded side of the barrier. In this work we present identical measurements carried out on four devices, but qualitatively matching behavior has been observed in about ten devices. For all measured devices, the maximum EBIC current is on the order of a few nA and the spatial peak-to-peak distances on the order of 100 nm. A slight asymmetry in peak height of the EBIC signal on each side of the barrier can be seen in most samples, but without any significant correlation to the wire's growth direction, the electric grounding point, or the symmetry of the barrier position relative to the contacts.

## 4. Discussion

The observed switch in current polarity around the location of the InP segment in the center of the nanowire (figure 2), along with the subsequent decay in current as the excitation source moves further away, is consistent with the mechanism for current generation proposed in figure 1(a). According to this model, the electron beam excites hot electron-hole pairs at a given location, with electrons expected to receive a higher portion of the kinetic energy than the holes (due to the asymmetry in effective mass). Because of their higher kinetic energy, electrons that travel towards the barrier have a higher chance of surmounting it than the holes, leading to charge separation. This matches well with the observation that locating the source of excitation on the left (right) side of the barrier results in a net flow of electrons toward the right- (left-) hand side of the device.

Importantly, we observe no current generation when the electron beam is positioned near the contacts. We can thus rule out the possibility that the observed current is generated at the metal-semiconductor interface, for example due to a Schottky contact. As a reference, EBIC was performed on devices made from pure InAs (WZ) nanowires that contained no InP barrier, where no signal resembling that of figure 2 could be observed (see supplemental information for details).

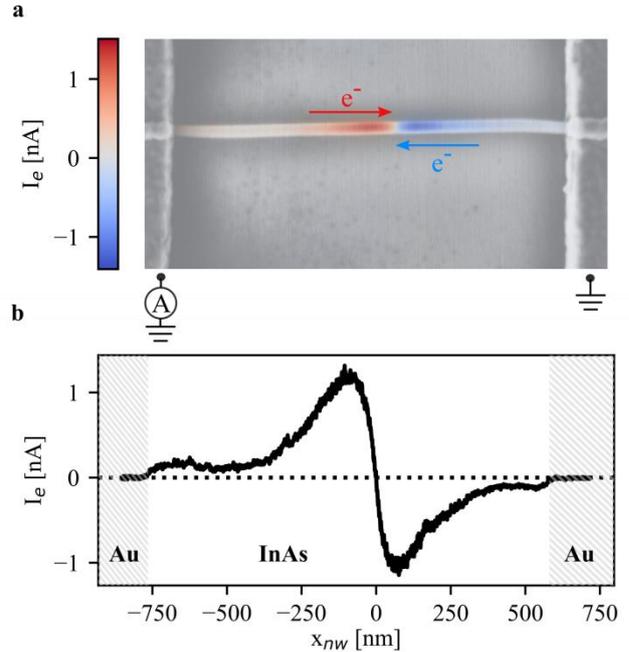

**Figure 2.** EBIC measurement on InAs/InP NW. (a) Composite image of SEM and EBIC data for a single NW device. The right metal contact is connected directly to ground, the left contact passes through an amplifier and Ampere meter setup. Positive (red) current is thus defined as electrons traveling towards the right (ground) in the image, as indicated by the red and blue arrows. (b) Line profile of EBIC current along the nanowire including Au contacts (grey shaded area).

The intuitive model for the current generation is based on the notion that hot carriers gradually lose energy as they approach the barrier. In the following we show that our data are consistent with such a model. In an EBIC experiment, the actual energy of excited carriers is not known and may well have a large spread. The rate of energy loss will depend on the inelastic carrier-carrier and carrier-phonon scattering rates. Both of these rates, as well as electron-hole recombination, are processes that decay exponentially and can be characterized with a diffusion length. In our model, we therefore describe energy decay by a single, effective electron diffusion length $L_e$.

Due to the higher effective mass of holes in InAs [14], we assume that the majority of the excitation energy go towards electrons, such that electrons are the majority charge carrier in the resulting EBIC. For the initial distribution of excited electrons $G_e(x)$, we use a Gaussian curve centered on the location of the electron beam, $x_e$.

$$G_e(x) = Ae^{-\frac{(x-x_e)^2}{2w^2}} \quad (1)$$

Here, $A$ is a normalization constant that is chosen such that the highest current value of data and model are aligned. For the root mean square width we use $w = 60$ nm, based on Monte Carlo simulations of the electron beam excitation volume with CASINO v2.5.1 [25] (see suplemental information). As the origin of $x_e$ and $x$ we choose the point where the EBIC current reverses direction, presumed to be the location of the InP barrier (expected from TEM images prior to device fabrication, see figure 1 (b)).

Based on the initial distribution, we assume a probability for electrons to cross the barrier that decreases exponentially with their distance from the barrier, $x$. The net current, $I_e$, generated by the electron beam at location $x_e$, is then proportional to the difference in the flow of electrons crossing the barrier from the left and right side, respectively.

$$I_e(x_e) \sim \int_{\text{Left contact}}^{0} G_e(x) e^{-x/L_e} dx - \int_{0}^{\text{Right contact}} G_e(x) e^{-x/L_e} dx \quad (2)$$

Figure 3 shows the EBIC current along four different nanowires together with fits of equation 2 and the fit value $L_e$ (see supplemental information for fitting procedure). The overall good quality of the fit supports the model based on figure 1 (a) and equation 2. Averaged over the four presented measurements (see supplemental information), we find $L_e = 110$ (±30) nm.

This value, found here for electrons, agrees well with an effective relaxation length for holes of about 100 nm in InAs found in a similar EBIC study of an axial InSb/InAs heterojunction [22]. Such a junction results in a type-III broken-band alignment, where minority carrier holes are extracted from the InAs to the InSb segment. However, it is likely that the energy relaxation length depends on the excitation method and on the initial energy distribution of hot carriers. For example, a recent study in which plasmonic elements were used to optically excite electron-hole pairs (photon energy between 1 and 1.3 eV), found an effective relaxation length on the order of 30 - 40 nm for electrons in InAs. [26] Another study electrically launched hot electrons with kinetic energies of about 500 meV in InAs nanowires and observed ballistic transport with a mean free path of 200-260 nm. [27]

An EBIC signal qualitatively similar to that of figure 2 and figure 3 has previously been observed in InAs nanowires containing an axial double InP-barrier [21]. We believe that a similar mechanism as described in figure 1a and equation 2 may have played a role also in that case.

## 5. Conclusion and Outlook

The presented results, enabled by the high spatial resolution of an electron beam, support the concept that a potential barrier embedded in a nanowire can be used to separate hot charge carriers. The results support the model illustrated in figure 1(a) and the interpretation of previous observations of photocurrent generation when the device was globally illuminated by optical light[10,11].

The $L_e \approx 100$ nm extracted here for electrons in InAs, and in Ref. [16] for holes, serves as a quantitative guide for the region available for effective hot-carrier extraction. The model used to interpret the data assumes transport in only one dimension (along the nanowire), employs a single effective relaxation length, and contains no information about the excitation energy. For this reason it will be valuable to repeat similar studies using optical excitation. Using optical light as an excitation source will allow for high spectral resolution and probing using different energy regimes, as well as for simultaneous measurement of power generation and efficiency. In such a study, we envision the study of energy regimes where transport across the barrier is dominated either by hot carriers that are transmitted ballistically (internal photoemission), or by hot carriers that have thermalised amongst each other (photo-thermionic emission). For sufficient control over the excitation source we envision the use of various methods such as optical beam-induced current (OBIC), scanning nearfield optical microscopy (SNOM), and the use of plasmonic elements along the nanowire to focus the absorption of light in small regions. It will then also be interesting to explore the role of barier height, thickness and geometry in effectively separating carriers.

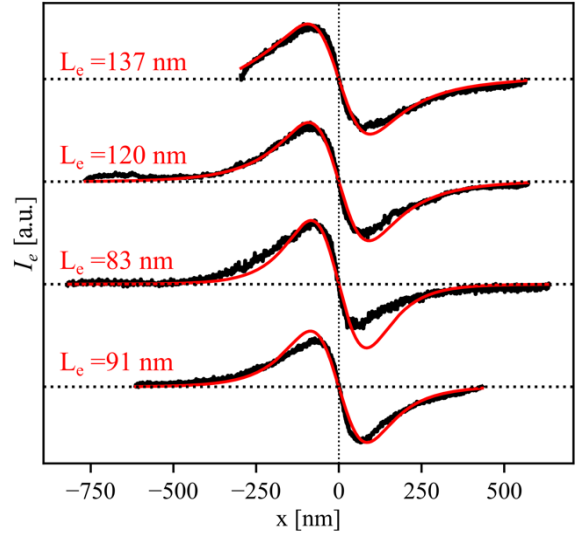

**Figure 3.** Normalized line profiles of EBIC data along four different nanowires (blue) with corresponding fit (red) and effective electron diffusion length, $L_e$.

## Acknowledgements

The authors acknowledge financial support by NanoLund, by the Swedish Energy Agency, by the Knut and Alice Walenberg Foundation (KAW) (project 2016.0089), the Swedish Research Council (VR) (project 2015-00619), the Marie Skłodowska Curie Actions, Cofund, Project INCA 600398.

# Hot-Carrier Separation in Heterostructure Nanowires observed by Electron-Beam Induced Current

Jonatan Fast,[1] Enrique Barrigon,[1] Mukesh Kumar,[1] Yang Chen,[1] Lars Samuelson,[1] Magnus Bogström,[1] Anders Gustafsson,[1] Steven Limpert,[1] Adam Burke,[1] Heiner Linke[1]

[1] NanoLund and Solid State Physics, Lund University, Box 118, 22100 Lund, Sweden

Supplemental information:

**Filtering of EBIC signal**

In this section, we describe the treatment that was done to the data in order to reduce noise in the form of horizontal lines that can be seen in the original data (figure S1 (a)). These sort of lines have been observed in previous EBIC measurements on single nanowires with similar device design, [1] and are believed to originate from charging of the insulating $SiO_2$ substrate as they always appear along the scan-direction of the electron-beam. In order to filter out parts of these charging effects, all the data are treated with a script that subtracts the median value of the EBIC along each horizontal line, resulting in the smoothened version seen in figure S1b. Finally, the SEM and the treated EBIC data are combined to form the composite image seen in figure S1c. Figure S2 shows the line profile of EBIC current of the same device, before and after the treatment.

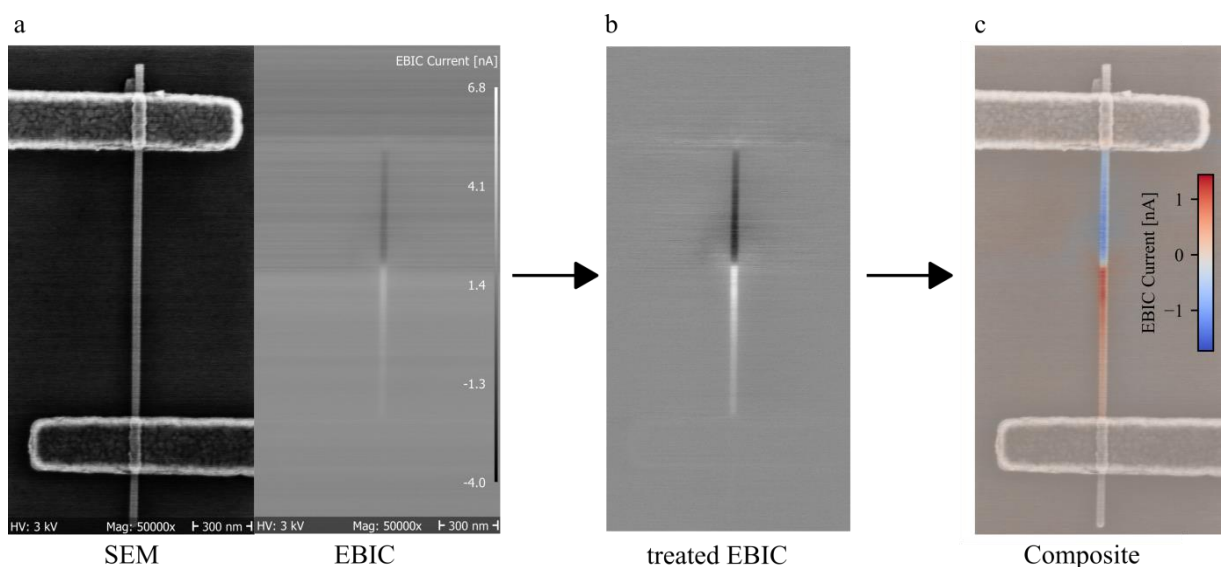

**Figure S1. (a)** SEM and EBIC data, as collected. **(b)** Treated EBIC data. **(c)** Composite of SEM and treated EBIC.

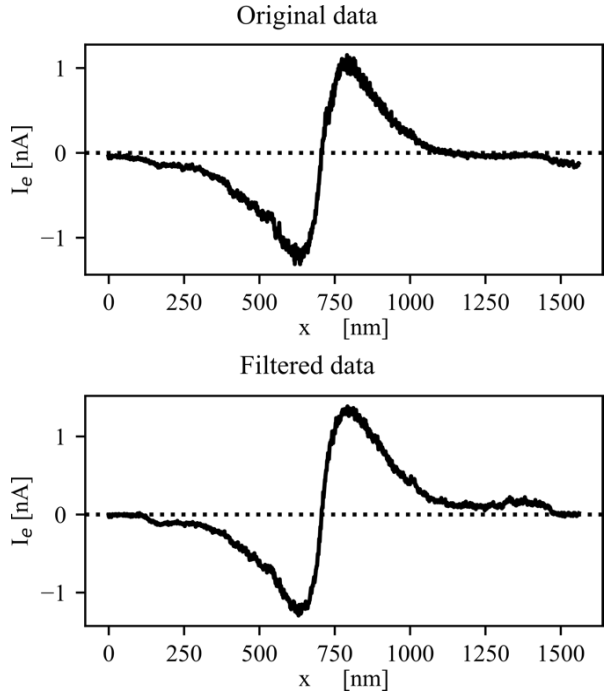

**Figure S2.** Line profile of EBIC data for same device as figure S1. Top/bottom without/with filtering of the data.

**Additional EBIC data**

Table S1 shows parameters extracted from the same EBIC data presented in figure 3 of the main text, the devices are numbered 1-4 corresponding to top-bottom in figure 3. The table contains the difference in current between the two peaks in each device, $I_{pp}$, the spatial separation between the peaks, $x_{pp}$, and the estimates of the effective electron diffusion length $L_e$. For each parameter, the mean value and standard deviation is calculated.

| Device # | $I_{pp}$ [nA] | $x_{pp}$ [nm] | $L_e$ [nm] |
|---|---|---|---|
| 1 | 2.63 | 154 | **137** |
| 2 | 2.5 | 157 | **120** |
| 3 | 0.98 | 134 | **83** |
| 4 | 2.1 | 145 | **91** |
| **Mean:** | **2.1 (±0.7)** | **150(±10)** | **110(±30)** |

**Table S1.** The current difference, $I_{pp}$, and spatial separation, $x_{pp}$, from peak-to-peak in the data, as well as for the estimated $L_e$, for each device is presented in a table with corresponding mean values and standard deviation.

**Comparison with pure InAs nanowires**

As a control, to check for the role of the InP energy barrier, EBIC was performed on devices with single InAs (WZ) nanowires (containing no InP barrier), fabricated in an identical way as described in the main text. As discussed below, the EBIC signals from such plain nanowires were qualitatively very different from those containing a barrier, supporting that the signal presented in the main text is indeed related to the energy barrier in the wire.

The result of one such measurement is seen in figure S3 a),b). A growing EBIC signal is observed the closer excitation gets to the contacts. Qualitatively, we find similar data for all tested devices, with the maximum of the observed EBIC current varying in the range of 0.1 to 1 nA. No strong signal is seen around the center of the wire, in contrast to the case of nanowires containing a barrier. Crucially, the polarity of the observed current is in the direct opposite to the observed current for devices with a barrier: here, the electron beam introduces a net flow of electrons towards the nearest contact.

The origin of the signal seen in the nanowire without barrier may be related to the nature of the metal – n-type InAs interface at the contacts, where the Fermi level is well-known to be pinned above the conduction band at the surface. [2] This results in a downwards band-bending as seen in figure S3 c). This band-bending allows for ohmic transport of electrons across the junction, but presents a barrier for holes in the valence band and thus might hinder the transport of holes. In this way, we think electron transport across the junction might be favored over hole transport, leading to a net flow of electrons towards the closest contact.

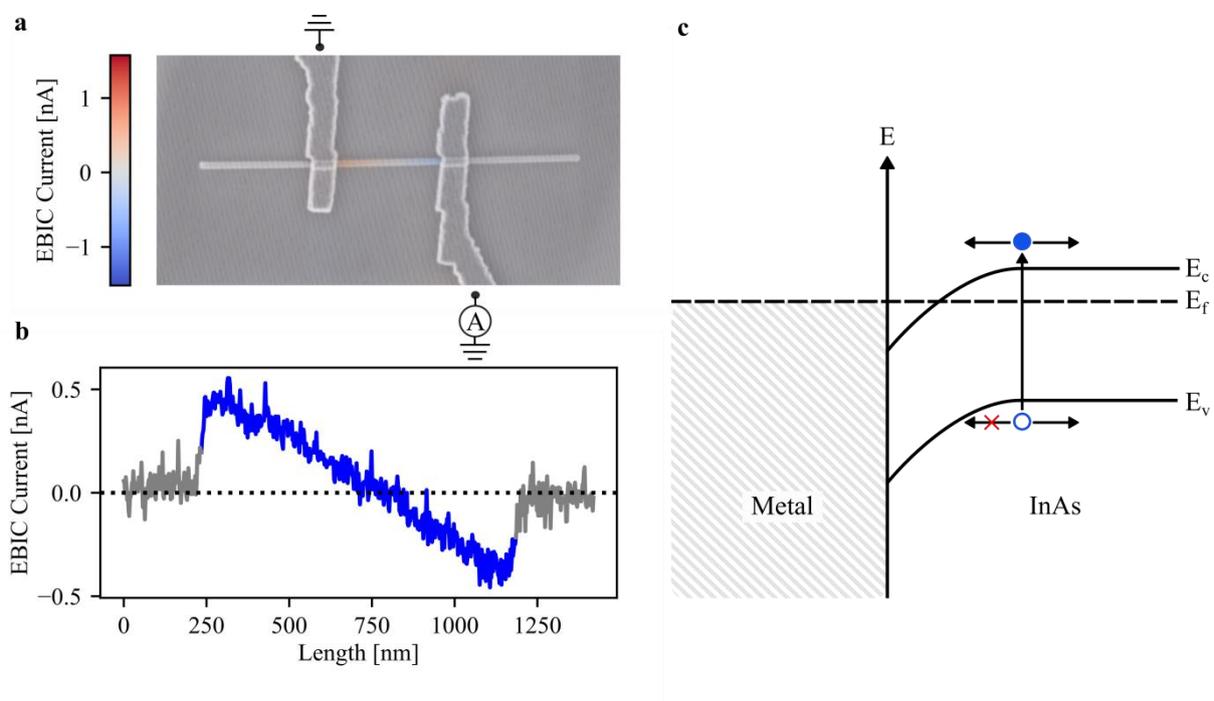

**Figure S3.** EBIC measurement on an InAs NW. a) Composite image of SEM and EBIC data for a single NW device. b) Line profile of EBIC current along the nanowire. c) Sketch of the expected band-bending at a metal – n-type InAs interface.

However, regardless of the origin of the signal, we expect that this signal is suppressed in wires containing an energy barrier, and therefore does not play an important role for the discussion in the main paper. We base this argument on current-voltage characteristics of nanowires with and without barrier (figure S4). In the nanowire without a barrier, we observe a resistance of about 6 kΩ (figure S4 (a)). Assuming a current on the order of 1 nA (figure S3), Ohm law tells us that the observed EBIC would correspond to a drift current generated by a voltage of around 6 µV. This is much smaller than the approximately 0.3-0.4 V that is needed to drive current across the barrier (figure S4 (b)).

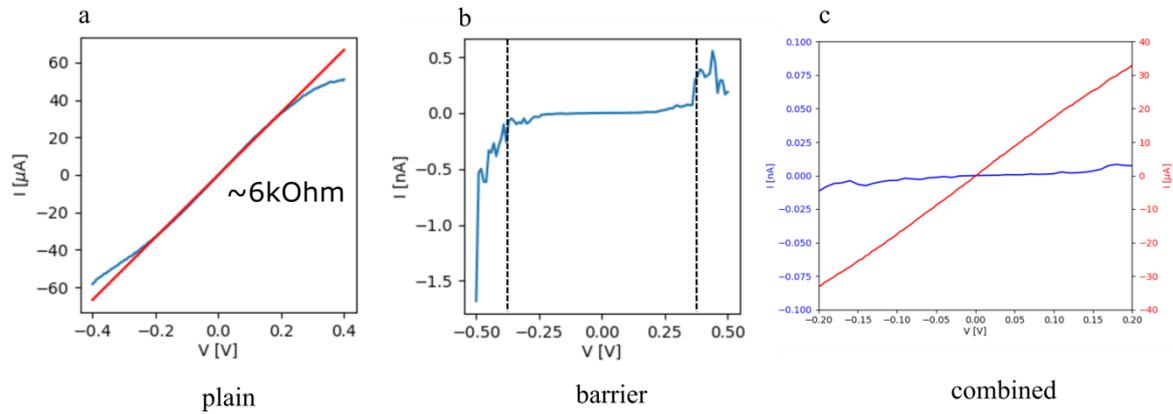

| plain | barrier | combined |

**Figure S4.** Current voltage characteristics of a) InAs nanowire with no barrier (same device as figure S3). b) InAs nanowire with InP barrier (same device as figure 2 of main text). c) plotted in the same window but with different scale.

**Excitation volume**

The excitation volume formed in an InAs substrate by an electron beam of 3kV was calculated using the software CASINO (monte CArlo SImulation of electroN trajectory in sOlids). [3] The software simulates the trajectory of electrons in a solid as a series of single scattering Monte Carlo events. The contour lines of figure S4 shows in which regions energy is deposited, the percentage represents how much of the incoming energy is deposited outside of the respective countour line. As shown in figure S5, for our models we choose an excitation volume of 60 nm, corresponding to the within which 95% of the electron beam energy is deposited. This value serves as the root mean square width, $w$, in equation 1 of the main text.

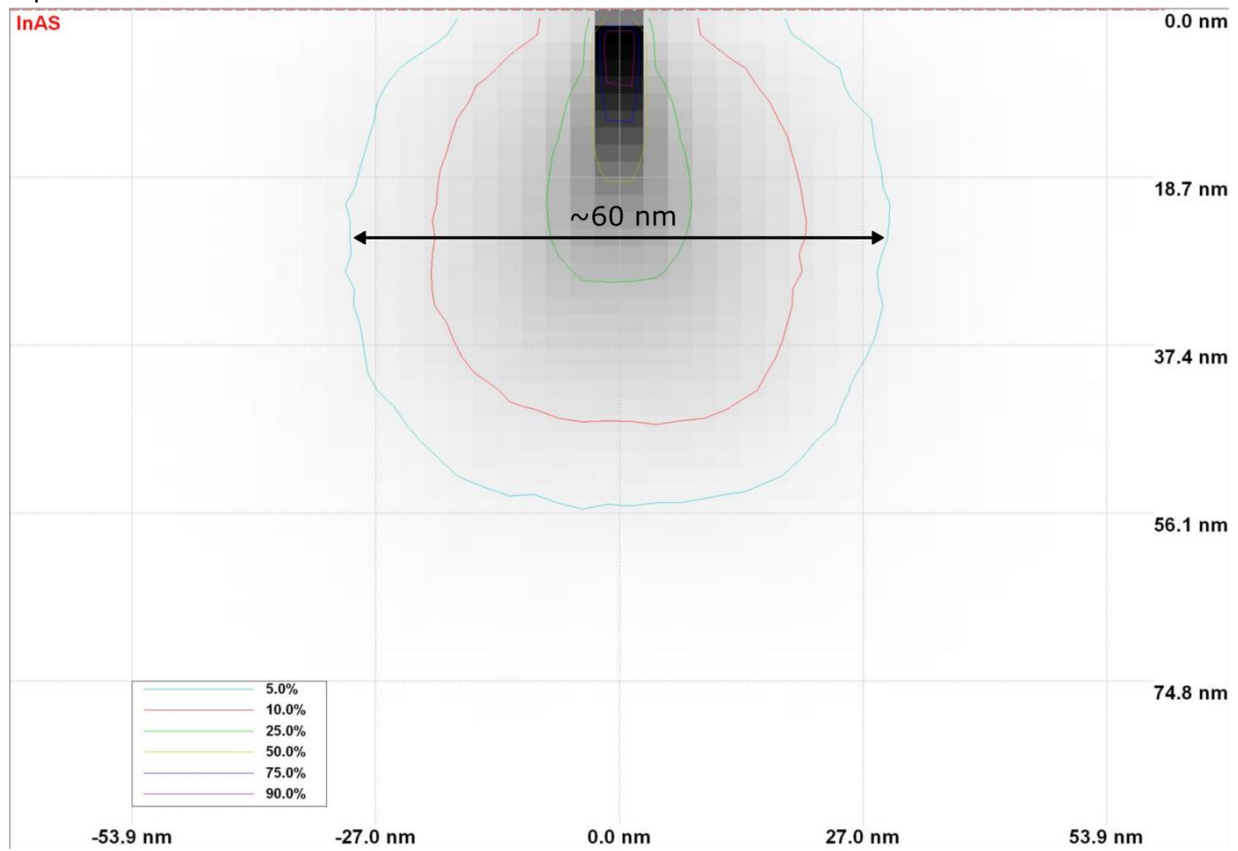

**Figure S5.** Monte Carlo modelling of energy deposited by an electron beam with 3kV in InAs. The largest region with a width of ~60 nm represents the region within which 95% of the beams energy is deposited.

**Fitting procedure**

In this section we explain how equation 2 of the main text was fitted to the EBIC data. The only unknown parameter in equation 2 is the effective diffusion length, $L_e$, and the fitting is performed by finding the value $L_e$ that minimizes the residual sum of squares (RSS) between the data and the model:

$$RSS = \sum_i^n (I_{data}(x_i) - I_{model}(x_i))^2 \quad (\text{Eq. S1})$$